\documentstyle[11pt,epsfig,astron]{article}

\begin{document}

\begin{titlepage}
\title{Distances in inhomogeneous quintessence cosmology}
\author{{M. Sereno, G. Covone, E. Piedipalumbo, R. de Ritis}\\
{\em \small $^1$ Dipartimento di Scienze Fisiche, Universit\`{a} di Napoli,}\\
{\em\small$^2$Istituto Nazionale di Fisica Nucleare, Sezione di Napoli,}\\
{\em \small Complesso Universitario di Monte S. Angelo, Via Cinzia,
Edificio G 80126 Napoli, Italy;}}.
\date{}
\maketitle
\begin{abstract}
We investigate the properties of cosmological distances in locally
inhomogeneous universes with pressureless matter and dark
energy ({\em quintessence}), with constant equation of state,
$p_X=w_X\rho_X$, $-1
\leq w_X <0$. We give exact solutions for angular diameter distances in the
empty beam approximation. In this hypothesis, the distance-redshift
equation is derived fron the multiple lens-plane theory. The case of a flat
universe is considered with particular attention. We show how this general
scheme makes distances degenerate with respect to $w_X$ and the smoothness
parameters $\alpha_i$, accounting for the homogeneously distributed
fraction of energy of the $i-$components. We analyse how this degeneracy
influences the critical redshift where the angular diameter distance takes
its maximum, and put in evidence future prospects for measuring the
smoothness parameter of the pressureless matter, $\alpha_M$.
\end{abstract}
\vspace{20.mm}
e-mail
addresses:\\sereno@na.astro.it\\covone@na.infn.it\\ester@na.infn.it\\
\vfill
\end{titlepage}
\section{Introduction}
Recent measurements of the cosmic microwave background anisotropy
\cite{deb&al00} strongly support the flat universe predicted by
inflationary cosmology. At the same time, there are strong evidences, i.e.
from dynamical estimates or X-ray and lensing observations of clusters of
galaxies, that the today pressureless matter parameter of the universe
$\Omega_M$ is significantly less than unity ($\Omega_M =\Omega_{CDM}
+\Omega_{B}
\simeq [0.30 {\pm} 0.10] +[0.04 {\pm} 0.01]$, where CDM and B respectively stand
for cold dark matter and baryons), while the hot matter (photons and
neutrinos) is really neglectable. Flatness so requires another positive
contribute to the energy density of the universe, the so called dark energy
-- see, for example, Silveira \& Waga \cite{si&wa97} or Turner \& White
\cite{tu&wh97} -- that, according to the recent data from SNIa
\cite{ri&al98,pe&al99}, must have negative pressure to account for an
accelerated expanding universe. One of the possible solutions of this
puzzle is the $\Lambda$CDM universe where the fraction $\Omega_X$ of the
dark energy is supported by a cosmological constant $\Lambda$. A very
promising altervative, the quintessence, was proposed by Caldwell, Dave \&
Steinhardt \cite{ca&al98} in the form of a dynamical, spatially
inhomogeneous energy with negative pressure, i.e. the energy of a slowly
evolving scalar field with positive potential energy. Apart from flatness,
a scalar field coupled to matter through gravitation can also explain the
so called coincidence problem, that is, it can provide a mechanism to make
the today densities of matter and dark energy comparable. As shown in Wang
et al. \cite{wa&al00}, a large set of indipendent observations agrees with
a universe described in terms of pressureless matter and quintessence
only, a hypothesis that we will make in this paper.

A scalar field is not an ideal adiabatic fluid \cite{gr94,ca&al98} and the
sound speed in it varies with the wavelength in such a way that high
frequency modes remain stable still when $w_X < 0$. Moreover, smoothness
is gauge dependent, and so a fluctuating inhomogeneous energy component is
naturally defined \cite{ca&al98}. Inhomogeneities, both in quintessence
and CDM, make the relations for the distances derived in
Friedmann-Lema\^{\i}tre-Robertson-Walker (FLRW) models not immediately
applicable to the interpetration of experimental data both in measurements
of luminosity distances and angular diameter distances. The observed
universe appears to be homogeneously distributed only on large scales
($\stackrel{>}{\sim} 500$ Mpc), while the propagation of light is a local
phenomena. In lack of a really satisfactory exact solution for
inhomogeneous universes in the framework of General Relativity
\cite{kr97}, the usual, very simple framework we shall adopt for the study
of distances is the {\em on average FLRW universe} (Schneider, Ehlers \&
Falco 1992; Seitz, Schneider
\& Ehlers 1994), where: {\em i)} the relations on a large scale are the
same of the corresponding FLRW universe; {\em ii)} the anisotropic
distortion of the bundle of light rays contributed by external
inhomogeneities (the shear $\sigma$) is not significant; {\em iii)} only
the fraction $\alpha_i$, the so called smoothness parameter, of the
$i$-component contributes to the Ricci focusing ${\cal R}$, that is to the
isotropic focusing of the bundle. The distance recovered in this 'empty
beam approximation', sometimes known as Dyer-Roeder (DR) distance, has been
long studied (Zel'dovich 1964; Kantowski 1969; Dyer \& Roeder 1972, 1973;
Linder 1988) and now is becoming established as a very useful tool for the
interpretation of experimental data (Kantowski 1998; Kantowski \& Thomas
2000; Perlmutter et al. 1999; Giovi, Occhionero
\& Amendola 2000). For a very large sample of standard candles, as SNIa for
luminosity distance, or standard rods, as compact radio sources for angular
diameter distances (Gurvits, Kellermann
\& Frey 1999), the effects of overdensity and underdensity along different
lines of sight between observer and sources balance each other out, and the
mean distance of the probability distribution of distances for fixed source
redshift, corresponding to the FLRW distance, can be used (Wang 1999,
2000). Nevertheless, for the present quite poor samples, sources are very
rarely brightened by the gravitational effect of occurring overdense
clumps, and so, for propagation of light far from local inhomogeneities,
the source appears dimmer and smaller with respect to the homogeneous case.
So the distance to be used is the mode of the distribution, that is, in a
very good approximation, the DR distance \cite{sef92,kan98}. In general,
the smoothness parameters for matter and quintessence (respectively
$\alpha_M$ and $\alpha_X$) and the equation of state $w_X$ are redshift
dependent, but they can be considered, in the redshift interval covered by
observations, as constant. For the clumpiness parameters, the constancy is
motivated by the absence of significant variations in the development of
structures in the observed redshift range. Instead, also in presence of
redshift variations in $w_X$, only its integral properties have effect on
the distance. In fact, in flat FLRW models the distance depends on $w_X$
only through a triple integral on the redshift (Maor, Brustein \&
Steinhardt 2000). Hereafter, we will consider the $\alpha_i$ and $w_X$ as
constant. For the cosmological constant $w_X=-1$; a network of light,
nonintercommuting topological defects \cite{vi84,sp+pe97} gives $w_X =-m/3$
where $m$ is the dimension of the defect: for a string $w_X =-1/3$, for a
domain wall $w_X=-2/3$; an accelerated universe requires $w_X <-1/3$.

In this approximation, distances are functions of a family of parameters:
$\Omega_M$ and $\Omega_X$, which describe the energy content of the
universe on large scales; $w_X$, which describes the equation of state of
the quintessence and varies between $-1$ and $0$; two clumpiness
parameters, $\alpha_M$ and $\alpha_X$, representing the degree of
homogeneity of the universe and which are to be used for phenomena of local
propagation.

This paper is organized as follows: in Sect.2 we introduce the so called DR
equation and discuss some features and its solution for quintessence in the
form of topological defects; Sect.3 lists the general solution for the DR
equation in the flat case and some simple expressions in extreme
situations; in Sect.4 the multiple lens-plane theory of gravitational
lensing is applied for a derivation of the DR equation without using the
focusing equation; Sect.5 is devoted to the degeneracy of the distance with
respect to the various parameters; in Sect.6 the critical redshift at which
the angular diameter distance takes its maximum is studied, and, at last,
in Sect.7 we draw the conclusions.

\section{The beam equation for inhomogeneous quintessence}

In the hypotheses discussed above, the focusing equation \cite{sa61,sef92}
for the angular diameter distance $D_A$ in terms of an affine parameter
$\lambda$,
\begin{equation}
\label{dr1}
\frac{d^2 D_A}{d \lambda^2}=-(|\sigma(\lambda)|^2 -{\cal R}(\lambda))D_A,
\end{equation}
becomes (see also Linder 1988)
\begin{equation}
\label{dr2}
\frac{d^2 D_A}{d \lambda^2}+\frac{1}{2}(1+z)^2\left[ 3\alpha_M\Omega_M
(1+z)^3+ n_X \alpha_X \Omega_X (1+z)^{n_X} \right]D_A=0,
\end{equation}
where $n_X \equiv 3(w_X +1)$ ($0<n_x<3$), and the relation between
$\lambda$ and the redshift $z$, in terms of the generalized Hubble
parameter
\begin{equation}
\label{dr3}
H(z)=H_0\sqrt{\Omega_M(1+z)^3+\Omega_X(1+z)^{n_X}+\Omega_K(1+z)^2},\
\Omega_K
\equiv 1-\Omega_M-\Omega_X,
\end{equation}
is
\begin{equation}
\label{dr4}
\frac{dz}{d\lambda}=(1+z)^2\frac{H(z)}{H_0}.
\end{equation}
The isotropic focusing effect in equation (\ref{dr2}) is simply represented
by the multiplicative factor to $D_A$; this coefficient increases with
$\alpha_M$, $\alpha_X$ and $n_X$.

Equation (\ref{dr2}) is sometimes called the generalized DR equation.
Substituting for $\lambda$ in equation (\ref{dr2}) by using equation
(\ref{dr4}), we have
\begin{equation}
\label{dr5}
H^2(z)\frac{d^2D_A}{dz^2}+\left[ \frac{2H^2
(z)}{1+z}+\frac{1}{2}\frac{dH^2}{dz}\right]\frac{D_A}{dz} +
\frac{1}{2}(1+z)\left[ 3\alpha_M \Omega_M + n_X \alpha_X \Omega_X (1+z)^{n_X-3}
\right]D_A=0;
\end{equation}
the initial conditions on equation (\ref{dr5}) are
\begin{equation}
\label{dr6}
D_A(z_d,z_d)=0,
\end{equation}
\[
\left. \frac{d}{dz}D_A(z_d,z)\right|_{z=z_d}=\frac{1}{1+z_d}\frac{c}{H(z_d)},
\]
where $D_A(z_d,z)$ is the angular diameter distance between $z_d$ (that, in
general, can be different from zero, as occurs in gravitational lensing for
the deflector) and the source at $z$.

Changing to the expansion factor $a \equiv 1/(1+z)$, equation (\ref{dr5})
is
\begin{eqnarray}\label{dr7}
&&a^2\left[ \Omega_M +\Omega_X a^{3-n_X} +\Omega_K
a\right]\frac{d^2D_A}{da^2}-a\left[
\frac{3}{2}\Omega_M +\frac{n_X}{2}\Omega_X a^{3-n_X} +\Omega_K a\right]\frac{d D_A}{da}\nonumber\\
&& +\left[ \alpha_M \frac{3}{2}\Omega_M +\alpha_X \frac{n_X}{2}\Omega_X
a^{3-n_X} \right]D_A=0,
\end{eqnarray}
a form which will be useful in the next sections.

In a generic space-time, the angular diameter distance $D_A$ and the
luminosity distance $D_L$ are related by (Etherington 1933)
\begin{equation}
\label{dr8}
D_L=(1+z)^2D_A,
\end{equation}
so that the considerations we will make about $D_A$ are easily extended to
$D_L$.

\subsection{Exact solutions for $\Omega_K\neq 0$}

The observational data nowadays available are in agreement with the
hypothesis of a flat universe, but are also compatible with a non zero,
although small, value of $\Omega_K$. A small value of $\Omega_K$ is also
allowed by the inflationary theory. These circumstances make useful the
study of the effect of the curvature on the cosmological distances since
today technology allows to put strong constraints on the cosmological
parameters.

For $\alpha_X=1$, equation (\ref{dr7}) reduces to
\begin{eqnarray}
\label{ok1}
 &&a^2(\Omega_M+ \Omega_K a+\Omega_X a^{3-n_X})\frac{d^2D_A}{da^2}-
 a\left(\frac{3}{2} \Omega_M+ \Omega_K a + \frac{n_X}{2} \Omega_Xa^{3-n_X}\right)
 \frac{d D_A}{da}+\nonumber\\
 &&\left(\frac{3}{2}\alpha_M+\frac{n_X}{2} \Omega_Xa^{3-n_X}\right)D_A=0.
\end{eqnarray}
To solve equation (\ref{ok1}), we proceed as in Demiasnki et al.
\cite{de&al00}. First, we look for a solution in the power form $a^s$ when
$\Omega_X
=\Omega_K
=0$. The parameter $s$ is constrained to fulfill the algebraic equation
\begin{equation}
\label{gen3}
s^2 -\frac{5}{2}s+\frac{3}{2}\alpha_M=0,
\end{equation}
which has the solutions
\begin{equation}
\label{gen4}
s_{\pm} = \frac{5}{4} {\pm} \frac{1}{4}\sqrt{25-24\alpha_M} \equiv \frac{5}{4} {\pm}
\beta .
\end{equation}
When $\Omega_X \not= 0$, $\Omega_K \not= 0$, we choose to impose the form
$D_A= a^s f(a)$ to the solution, being $f$ a generic function. Inserting
this expression into equation (\ref{ok1}) we have for $f$
\begin{eqnarray}
&& a \left(\Omega_M +\Omega_K a+\Omega_X a^{3-n}
\right)\frac{d^2f}{da^2} - \frac{df}{da}\left(\Omega_M (2s-\frac{3}{2})+\Omega_K (2s-1)a
\right.\\&&+ \left. \left( s (s-1)- \frac{n_X}{2} s\right)\Omega_Xa^{3-n_X}\right)+(\Omega_K s(s-2)
+ \Omega_X (s^2-s (1+\frac{n_X}{2}))a^{2-n_X})f=0.\nonumber
\label{ok2}
\end{eqnarray}
The initial conditions at $a=1$ for the auxiliary function $f$ come out
from equation (\ref{dr6}) evaluated at $z=0$,
\begin{equation}
\label{ok3} f(1)=0,
\end{equation}
\[
\left. \frac{d}{dz}f(a)\right|_{a=1}=\frac{c}{H_0}.
\]

Equation (\ref{ok2}) is very useful to obtain some exact solutions of the
DR equation, corresponding to integer values of the quintessence parameter
$n_X$. Since the solutions for the case of the cosmological constant,
$n_X=0$ \cite{kan98}, and for the pressureless matter, $n_X=3$
\cite{se&sc94}, are already known, we will consider only string networks,
$n_X=2$ $(w_X=-1/3)$ and the domain walls with $n_X=1$ $ (w_X=-2/3)$. Let
us start with the case $n_X=2$, when equation (\ref{ok2}) reduces to
\begin{equation}
\label{ok4}
  a(c_1+ c_2 a)\frac{d^2 f}{d a^2} + (c_3+ c_2 c_4 a)\frac{d f}{d a}
  + c_5 f == 0,
\end{equation}
being
\begin{eqnarray}
\label{ok5}
c_1&=&\Omega_M, \nonumber \\ c_2&=&\Omega_K+\Omega_X, \nonumber \\
c_3&=&c_2\left(2s-\frac{3}{2}\right),\nonumber\\ c_4&=&2s-1,\nonumber \\
c_5&=&\Omega_K s(s- 2) +\Omega_X\left( s\left(s - \frac{3}{2} +
\frac{1}{2}\right) \right) .
\end{eqnarray}
Equation ~(\ref{ok4}) is of hypergeometric  type, i.e. it has three regular
singularities \cite{inc56}, and so, for $n_X =2$, $f$ is the hypergeometric
function. If we indicate with $f_{s_{+}}$ and $f_{s_{-}}$ two indipendent
solutions for, respectively, $s=s_{+}$ and $s
=s_{-}$, we can write the general solution of equation (\ref{ok1}) for
$n_X =2$ as
\begin{eqnarray}\label{ok7}
D_A&=&A_{+}a^{s_{+}}f_{s_{+}}(a) + A_{-}a^{s_{-}}f_{s_{-}}\left( a
\right)\nonumber\\ &=&\frac{1}{(1+z)^{5/4}}\left( A_+(1+z)^{-\beta}f_{s_{+}}
\left(\frac{1}{1+z}\right) + A_{-}\left( 1+z \right)^{\beta} f_{s_{-}}
\left( \frac{1}{1+z}\right)\right),
\end{eqnarray}
where $A_{+}$ and $A_{-}$ are constants determined by the initial
conditions. In equation (\ref{ok7}) we have expressed the scale factor,
$a$, in terms of the redshift.

Let us consider now the case $n_X=1$, when the equation for $f$ becomes
\begin{eqnarray}
\label{ok8}
\lefteqn{a\left(\frac{\Omega_M}{\Omega_X}+ \frac{\Omega_K}{\Omega_X} a+
a^{2}\right)\frac{d^2 f}{da^2}} \\&& +\left(\frac{
\Omega_M}{\Omega_X}\left(2 s-\frac{3}{2} \right)+ \frac{
\Omega_K}{\Omega_X}
\left(2s-1 \right)a + \left( 2s-\frac{1}{2} \right)a^{2} \right)\frac{d f}{da} \\
&& + \left(\frac{\Omega_K}{\Omega_X} s\left(s-2\right)+\left(s\left(
s-2\right)+\frac{1}{2}\right)a \right)f=0.
\end{eqnarray}
Equation (\ref{ok8}) is a fuchsian equation with three finite regular
points plus a regular singularity at $\infty$ \cite{inc56}. The regular
points in the finite part of the complex plane are
\begin{eqnarray}
\label{ok9}
a_1&=& 0,\nonumber\\ a_2&= &
\frac{-\Omega_K-\sqrt{\Omega_{K}^2-\Omega_M\Omega_X}}{2},\\
a_3&= & \frac{-\Omega_K+\sqrt{\Omega_{K}^2-\Omega_M\Omega_X}}{2}. \nonumber
\end{eqnarray}
The trasformation $y=a/a_2$ sends $a_2\rightarrow 1$ and $a_2\rightarrow
\zeta= \frac{a_3}{a_2}$. In terms of $y$, equation ~(\ref{ok8}) is
\begin{eqnarray}\label{ok10}
&& y \left(y-1\right)\left(y-\zeta \right)\frac{{d^2 f}{dy^2}}
+\left(\frac{\Omega_M}{\Omega_X}(2 s-\frac{3}{2})+
\frac{\Omega_K}{\Omega_X} ( 2s-1) a_{2} y \right) \nonumber\\&&\quad
+\left((2 s- \frac{1}{2}){a_2}^{2}{y}^{2}\right)\frac{d f}{dy}
+\left(\frac{\Omega_K}{a_2 \Omega_X} s(s-2)+
\left(s(s-2)+\frac{1}{2}\right)a_2 y\right)f=0,
\end{eqnarray}
which can be reduced to the standard form
\begin{equation}
\label{ok11}
\frac{d^2 f}{dy^2}+\left(\frac{\gamma}{y}+\frac{\delta}{y-1}+
\frac{\epsilon}{y-\zeta}\right)\frac{d f}{d y}+\left(\frac{\theta\lambda
y-q}{y\left(y-1\right)\left(y-\zeta\right)}\right)f=0,
\end{equation}
where
\begin{eqnarray}\label{ok12}
\gamma+\delta&=&\left(2s-\frac{1}{2}\right)a_2^2,\nonumber \\
\frac{\Omega_K}{\Omega_X}\left(2s-1 \right)a_2&=&\gamma\left(1+\zeta\right)+
\delta\zeta +\epsilon,\nonumber\\
\gamma \zeta&=&\frac{\Omega_M}{\Omega_X}\left(2s-\frac{3}{2}\right),\\
q&=&\frac{\Omega_K}{\Omega_M}s\left(s-2\right),\nonumber\\
\theta\lambda&=&s\left(s-\frac{3}{2} \right)+\frac{1}{2}.\nonumber
\end{eqnarray}
Equation (\ref{ok11}) is the Heun equation \cite{erd+al55}, which is
slightly more complicated then the hypergeometric equation, possessing four
points of regular singularity in the entire complex plain, rather than
three. The constant $q$ is the so called {\it accessory parameter}, whose
presence is due to the fact that a fuchsian equation is not completely
determined by the position of the singularities and the indices. The Heun
equation can be charaterized by a ${\cal P}$ symbol, and the solutions can
be expanded in series of hypergeometric functions. Thus, the solution of
the equation ~(\ref{ok1}) for $n_X=1$ can be formally written as equation
~(\ref{ok7}), once the functions $f_{s_{+}}$ and $f_{s_{-}}$ are
interpreted as Heun functions.

In Fig. \ref{fig1}, we plot the solutions found above.

\section{Exact solutions for $\Omega_K =0$}

The DR equation for a flat universe has already been solved considering the
limiting case of the cosmological constant \cite{kan98,ka&th00,de&al00}.
Here, in presence of generic quintessence, we propose the general solution
in terms of hypergeometric functions and, then, list particular solutions
in terms of elementary functions.

\subsection{General solution}
When $\Omega_K =0$, equation (\ref{dr7}) reduces to
\begin{eqnarray}\label{gen1}
&&\left(a^2(\Omega_M +(1-\Omega_M) a^{3-n_X})\right)\frac{d^2
D_A}{da^2}-a\left(
\frac{3}{2}\Omega_M +\frac{n_X}{2}(1-\Omega_M) a^{3-n_X}\right)\frac{dD_A}{da}
\nonumber\\
&&+  \left(\frac{3}{2}\alpha_M\Omega_M+\frac{n_X}{2}\alpha_X (1-\Omega_M)
a^{3-n_X}\right)D_A=0;
\end{eqnarray}
dividing equation (\ref{gen1}) by $\Omega_M$ and defining $\mu \equiv
\frac{1-\Omega_M}{\Omega_M}$, we have
\begin{equation}
\label{gen2}
a^2(1 +\mu a^{3-n_X})\frac{d^2 D_A}{da^2}-a\left( \frac{3}{2}
+\frac{n_X}{2}\mu a^{3-n_X}\right)\frac{dD_A}{da}+\left(
\frac{3}{2}\alpha_M +\frac{n_X}{2}\alpha_X \mu a^{3-n_X}\right)D_A=0.
\end{equation}
To solve equation (\ref{gen2}), we proceed as in Sect.2.1. First, we look
for a solution in the power form $a^s$ when $\mu =0$. The parameter $s$ is
constrained to fulfill equation (\ref{gen3}). When $\mu \not= 0$, we choose
to impose the form $D_A = a^sf(a)$ to the solution, where $f$ is generic.
Inserting this expression into equation (\ref{gen2}) and changing to $x
\equiv a^{3-n_X}$, we have for $f$
\begin{eqnarray}\label{gen5}
&&x(3-n_X)(1+\mu x)\frac{d^2 f}{dx^2}+ \left( \left( 2s-n_X+
\frac{1}{2}\right)+ \left( 2-\frac{3}{2}n_X +2s\right)\mu x \right)\frac{d
f}{dx}\nonumber \\ &&+\frac{\mu}{2}\left( s-\frac{3\alpha_M
-n_X\alpha_X}{3-n_X}\right)f=0.
\end{eqnarray}
Again, this equation can be solved in terms of hypergeometric functions.
Denoting with $f_{s_+}$ and $f_{s_-}$ two of such indipendent solutions
for, respectively, $s
=s_+$ and $s =s_-$, we can write the general solution of equation (\ref{gen1}) as
\begin{eqnarray}\label{gen6}
&&\lefteqn{D_A=A_+a^{s_+}f_{s_+}(x(a)) + A_-a^{s_-}f_{s_-}(x(a))}
\nonumber
\\ &&  \\&&=\frac{1}{(1+z)^{5/4}}\left( A_+(1+z)^{-\beta}f_{s_+} \left(
\frac{1}{(1+z)^{3-n_X}}\right) +  A_-(1+z)^{\beta}f_{s_-} \left(
\frac{1}{(1+z)^{3-n_X}}\right)\right),\nonumber
\end{eqnarray} where $A_+$ and $A_-$ are constants determined by the
initial conditions.

\subsection{Particular cases}

Once we have the general solution of equation (\ref{gen1}) in terms of
hypergeometric functions, we go now to list some expressions of the angular
diameter distance in terms of elementary functions in two extremal cases.

\subsubsection{Homogeneous universe}

In this case we have that $\alpha_M =\alpha_X=1$, so that the angular
diameter distance takes the form valid in a FLRW universe, that is
\begin{equation}
\label{par1}
D_A(z_d,z)=\frac{c}{H_0}\frac{1}{1+z}\int_{z_d}^z
\frac{dz^{'}}{\sqrt{\Omega_M(1+z^{'})^3+(1-\Omega_M)(1+z^{'})^{n_X}}}.
\end{equation}
This is the integral of the differential binomial
\begin{equation}
\label{par1bis}
x^\mu(a+bx^\nu)^\rho,
\end{equation}
where $x=1+z,\  a=\Omega_M,\  b=1-\Omega_M,\  \mu=-3/2,\ \nu=n_X-3$ and
$\rho=-1/2$. We can put this integral in rational form when
\begin{equation}
\label{par2}
n_X=\frac{3s-1}{s}, \ s \in {\cal Z} -\{ 0\},
\end{equation}
performing the substitutions (Picone \& Miranda 1943)
\[
t=\sqrt{\Omega_M +(1+\Omega_M)(1+z)^{n_X-3}}
\]
 when $s$ is even and
\[
t=\sqrt{\frac{\Omega_M +(1+\Omega_M)(1+z)^{n_X-3}}{(1+z)^{n_X-3}}}
\]
for odd $s$. Equation (\ref{par2}) includes all and only the rational
values of $n_X$ for which equation (\ref{par1}) can be solved in terms of
elementary functions. $n_X$ varies from $2$ ($w_X =-1/3$), when
quintessence evolves like curvature, to $4$ ($w_X =1/3$)({\em hot dark
matter}); for $s
\rightarrow {\pm}
\infty$, $n_X$ tends to $3$, giving ordinary pressureless matter. For
$n_X=2$ (see also Lima \& Alcaniz 2000) we get
\begin{equation}
\label{par3}
D_A(z_d,z)=\frac{c}{H_0}\frac{2}{(1+z)\sqrt{1-\Omega_M}} \left[ {\rm
Arctanh} \left(
\frac{\sqrt{1+\Omega_M z}}{\sqrt{1-\Omega_M}} \right) \right]_{z=z_d}^z;
\end{equation}
we note that with respect to the dynamical equations, a flat universe with
$n_X=2$ behaves like an open one with $\Omega_K =1-\Omega_M \neq 0$, but,
on the other hand, while quintessence contributes to the Ricci focusing, a
geometric term does not. For $n_X=4$, it is
\begin{equation}
\label{par5}
D_A(z_d,z)=\frac{c}{H_0}\frac{2}{\Omega_M(1+z)}\left[
\frac{\sqrt{1+z-\Omega_M z}}{(1+z)^{1/2}}\right]_{z}^{z=z_d}.
\end{equation}
Equation (\ref{par5}) holds in the past history of the universe at the
epoch of matter-radiation equality ($z_{eq} \sim 10^4$). Other solutions
with $2<n_X<4$ are easily found. Even if they can be physically interesting
when related to other behaviours of the scale factor, they cannot explain
the today observed accelerated universe. So, we will not mention them here.

\subsubsection{Totally clumpy universes}

We now study  very particular models of universe in which both matter and
quintessence are totally clumped, that is $\alpha_M =\alpha_X=0$. In this
case, the DR equation reduces to a first order equation and the expression
for the angular diameter distance becomes
\begin{equation}
\label{par7}
D_A(z_d,z)=\frac{c}{H_0}(1+z_d)\int_{z_d}^z\frac{dz^{'}}{(1+z^{'})^2
\sqrt{\Omega_M(1+z^{'})^3+ \Omega_X(1+z^{'})^{n_X}}}.
\end{equation}
When $n_X=0$ ($w_X=-1$) and $\alpha_M=0$, the DR equation becomes of the
first order indipendently of the values of $\Omega_X$ and $\alpha_X$, and
so the distance takes the form
\begin{equation}
\label{par7bis}
D_A(z_d,z)=\frac{c}{H_0}(1+z_d)\int_{z_d}^z\frac{dz^{'}}{(1+z^{'})^2\sqrt{\Omega_M(1+z^{'})^3
+ \Omega_X}}.
\end{equation}

Once again, in equation (\ref{par7}) there is the integral of a
differential binomial of the form given in equation (\ref{par1bis}), with,
this time, $a=
\Omega_M , \ b= 1-\Omega_M ,\
\mu =-7/2, \
\nu=n_X-3$ and $\rho =-1/2$. When $n_X$ is rational, all and only the solutions of
equation (\ref{par7}) in terms of elementary functions occur when
\begin{equation}
\label{par8}
n_X=\frac{3s-5}{s},\ s \in {\cal Z} -\{ 0\};
\end{equation}
for any such $s$ we can perform the same substitutions already described
for homogeneous universes in the previous subsection. Now, we have values
of $n_X <2$: for $s=2,\ 3,\ 4$, respectively, we find $n_X\ (w_X)=1/2\
(-5/6),\ 4/3\ (-5/9),\ 7/4\ (-5/12)$. For $n_X=1/2$, it is
\begin{equation}
\label{par9}
D_A(z_d,z)=\frac{c}{H_0}\frac{4(1+z_d)}{5(1-\Omega_M)}\left( \sqrt{\Omega_M
+
\frac{1-\Omega_M}{(1+z_d)^{5/2}}} -
\sqrt{\Omega_M + \frac{1-\Omega_M}{(1+z)^{5/2}}} \right);
\end{equation}
for $n_X=4/3$, we get
\begin{eqnarray}\label{par10}
D_A(z_d,z)&=&\frac{c}{H_0}(1+z_d)\left[ \frac{6}{5}\Omega_M
\sqrt{1-\Omega_M +\Omega_M(1+z)^{5/3}} \left(
(\Omega_M -1)^2 +\right.\right.\\&+&\left.\left.\frac{2}{3}(\Omega_M
-1)^2
\left( 1-\Omega_M +\Omega_M(1+z)^{5/3} \right) + \frac{1}{5} \left(
1-\Omega_M +\Omega_M(1+z)^{5/3} \right)^{2}\right)\right]_{z=z_d}^z
,\nonumber
\end{eqnarray}
and, for $n_X =7/4$,
\begin{eqnarray}\label{par11}
D_A(z)&=&\frac{c}{H_0}\frac{8}{5(1-\Omega_M )^2}\left(
\left(\frac{1}{3}-\Omega_M\right) - \left(\frac{1}{3}\left(
\frac{1-\Omega_M}{(1+z)^{5/4}}+\Omega_M \right)^{3/2}\right.\right.\\&-&\left. \left.\Omega_M \sqrt{ \frac{1-
\Omega_M}{(1+z)^{5/4}}+\Omega_M}
\right) \right) .
\end{eqnarray}
Other interesting results are obtained when $n_X =2\ (s=5)$ and $n_X =4\
(s=-5)$. For $n_X =2$ (string networks), the angular diameter distance is
\begin{eqnarray}\label{par12}
D_A(z_d,z)&=&2\frac{c}{H_0}(1+z_d)\Omega_M^2 {\cal E}\left[ (\Omega_M-1)^3
+ (\Omega_M-1)^2{\cal E}\right. \nonumber\\  & +&\frac{3}{5}(\Omega_M
-1){\cal E}^2+\left.\frac{1}{7}{\cal E}^3  \right]_{z=z_d}^z,
\end{eqnarray}
where
\[
{\cal E}=\sqrt{1-\Omega_M +(1+z)\Omega_M};
\]
for $n_X =4$ (hot dark matter), it is
\begin{equation}
\label{par13}
D_A(z_d,z)=\frac{c}{H_0}\frac{1+z_d}{\Omega_M^3(\Omega_M -1)}\left[
\frac{{\rm Arctan}\left(
\sqrt{\frac{1+z-\Omega_M z}{(1+z)(\Omega_M -1)}}\right)}{\sqrt{1-\Omega_M}}
+\sqrt{\frac{(1+z)(1+z-z\Omega_M)}{\Omega_M}}
\right]^{z=z_d}_z .
\end{equation}
In the limit $s \rightarrow {\pm} \infty$, $n_X$ goes to 3 (CDM).

\section{An alternative derivation of the generalized DR equation}

As already shown for a universe with pressureless matter (Schneider \&
Weiss 1988; Schneider et al. 1992), it is possible to derive the DR
equation from the multiple lens-plane theory, without referring to the
focusing equation. We want, in the framework of the on average FLRW
universes, to generalize this result to the case of inhomogeneous
quintessence. The basic idea is the simulation of the clumpiness by adding
to a smooth homogeneous background a hypothetical density distribution of
zero total mass, which is made of two components: a distribution of clumps
(both in dust and dark energy) and a uniform negative energy density such
that the mean density of the sum of both components is zero. After such
addition, the average properties of the universe on large scales are still
that corrensponding to the background FLRW model. The gravitational surface
density $\Sigma$ of clumps in a shell of size $\Delta z$ centered on the
observer is then
\begin{equation}
\label{vol1}
\Sigma =\Delta z \frac{d r_{prop}}{dz}T^{00}_{cl},
\end{equation}
where the relation between the redshift and the proper distance $r_{prop}$
is that valid in FLRW universes,
\begin{equation}
\label{vol2}
\frac{d r_{prop}}{dz}=\frac{c}{H(z)}\frac{1}{1+z},
\end{equation}
and $T^{00}_{cl}$ is the 0-0 element of the total energy-momentum tensor in
clumps,
\begin{equation}
\label{vol3}
T^{00}_{cl}(z)=(1-\alpha_M)T^{00}_{M}+(1-\alpha_X)T^{00}_{X}=
(1-\alpha_M)\rho_M(z)c^2+(1-\alpha_X)(1+w_X)
\rho_X(z)c^2.
\end{equation}
In an on average FLRW universe, the densities of pressureless matter and
quintessence are, respectively,
\begin{equation}
\rho_M(z)=(1+z)^3\Omega_M \rho_{cr},
\end{equation}
\[
\rho_X(z)=(1+z)^{n_X} \Omega_X \rho_{cr},
\]
where $\rho_{cr} \equiv 3H_0^2/(8\pi G)$ is the today critical density. The
dimensionless surface density $k$ corresponding to equation (\ref{vol1}) is
\begin{eqnarray}\label{vol5}
k \Delta z &\equiv& \frac{4 \pi
G}{c^2}\frac{D_1(z)D_1(z,z_s)}{D_1(z_s)}\Sigma =
\frac{H_0^2}{cH(z)}\frac{(1+z)^3}{2}\left[3(1-\alpha_M)\Omega_M +\right.\nonumber\\
&&\quad + \left. n_X(1-\alpha_X)\Omega_X (1+z)^{n_X-3} \right]
\frac{D_1(z)D_1(z,z_s)}{D_1(z_s)}\Delta z,
\end{eqnarray}
where the subscript 1 refers to diameter angular distances in FLRW
universes and $z_s$ is a hypothetical source redshift. The so constructed
spherical shells will act as multiple lens-plane. The ray-trace equation
which describes successive deflections caused by a series of lens planes
is \cite{sef92}
\begin{equation}
\label{vol6}
{\bf x_j}={\bf x_1}-\sum
_{i=1}^{j-1}\frac{D_1(z_i,z_j)}{D_1(z_j)}{\mbox{\boldmath
$\hat{\alpha}_i$}},
\end{equation}
where ${\bf x_i}$ is the bidimensional angular position vector in the $i$th
lens plane and ${\mbox{\boldmath $\hat{\alpha}_i$}}$ is the deflection
angle a light ray undergoes if it traverses the $i$th lens plane at ${\bf
x_i}$. The lens planes are ordered such that $z_i <z_j$ if $i<j$. The solid
angle distortion is described by the $2{\times} 2$ Jacobianes matrices of the
mapping equation (\ref{vol6}),
\begin{equation}
\label{vol7}
A_i \equiv \frac{\partial {\bf x_i}}{\partial {\bf x_1}},
\end{equation}
and by the derivatives of the scaled deflection angle ${\mbox{\boldmath
${\alpha}_i$}}=(D_1(z_i,z_s)/D_1(z_s)){\mbox{\boldmath $\hat{\alpha}_i$}}$,
\begin{equation}
\label{vol8}
U_i \equiv \frac{\partial {\mbox{\boldmath ${\alpha}_i$}}}{\partial {\bf
x_i}}.
\end{equation}
By taking the derivative of equation (\ref{vol6}) with respect to the
indipendent variable ${\bf x_1}$, which represents the angular position of
an image on the observer sky, we have the recursion relation
\begin{equation}
\label{vol9}
A_j= {\cal I}- \sum
_{i=1}^{j-1}\frac{D_1(z_i,z_j)D_1(z_s)}{D_1(z_j)D_1(z_i,z_s)}U_i A_i,
\end{equation}
with $A_1={\cal I}$, ${\cal I}$ being the two-dimensional identity matrix.
In our model of a clumpy universe, the matrices $U_i$ are given by
\cite{sc&we88}
\begin{equation}
\label{vol10}
U_i = -k_i {\cal I}\Delta z  - {\cal T}_i ,
\end{equation}
where the first term accounts for the negative convergence caused by the
smooth negative surface density and ${\cal T}_i$ is the matrix that
describes the deflection caused by the clumps. In the empty beam
approximation (light rays propagating far away from clumps and vanishing
shear), it is ${\cal T}_i=0$ and then all the $A_i$ are diagonal,
$A_i=a_i{\cal I}$. Equation (\ref{vol9}) becomes
\begin{equation}
\label{vol11}
a_j= 1 + \sum
_{i=1}^{j-1}\frac{D_1(z_i,z_j)D_1(z_s)}{D_1(z_j)D_1(z_i,z_s)}k_i  a_i \Delta z,
\end{equation}
where the dependence on $z_s$ drops out in the product of the ratio of
distances by $k_i$ . In the continuum limit, $\Delta z
\rightarrow 0$, equation (\ref{vol11}) is
\begin{equation}
\label{vol12}
a(z)= 1 + \int _{0}^{z}\frac{D_1(y,z)D_1(z_s)}{D_1(z)D_1(y,z_s)}k(y)a(y)dy.
\end{equation}
Multiplying equation (\ref{vol11}) by $D_1(z)$ and letting
$D_A(z)=a(z)D_1(z)$, we obtain, substituting for the explicit expression of
$k$ given in equation (\ref{vol5}),
\begin{eqnarray}\label{vol13}
&&D(z)=D_1(z)+\left(\frac{H_0^2}{c}\right)\int_0 ^z\frac{(1+y)^2}{H(y)}
\left[\frac{3}{2}(1-\alpha_M)\Omega_M +\right.\nonumber\\&&
\left.\frac{n_X}{2}(1-\alpha_X)\Omega_X (1+y)^{n_X-3}
\right]D_1(y,z)D_A(z)dy.\nonumber
\end{eqnarray}
It is easy to verify that equation (\ref{vol13}) is equivalent to the
generalized DR equation equation (\ref{dr5}) with initial condition given
for $z_d =0$. Changing to $z_d$ the lower limit of the integration in
equation (\ref{vol13}) and $D_A(z)$ $(D_1(z))$ with $D_A(z_d,z)$
$(D_1(z_d,z))$, we have the equation for generic initial conditions.
Equation (\ref{vol13}), already derived with a different way of proceeding
by Linder \cite{li88}, has here been found only using the multiple
lens-plane theory.

Equation (\ref{vol13}) is a Volterra integral equation of the second kind
\cite{tr85} whose solution is
\begin{equation}
\label{vol14}
D_A(z)=D_1(z)+\int_0^z H(y,z)D_1(y)dy,
\end{equation}
where the resolvent kernel $H(y,z)$ is given by the series of iterated
kernels
\begin{equation}
\label{vol15}
H(y,z)=\sum_{i=0}^{\infty}K_i(y,z),
\end{equation}
with
\begin{eqnarray}\label{vol17}
K_1(y,z)= \left\{
   \begin{array}{cl}\left(
\frac{H_0^2}{c}\right)\frac{(1+y)^2}{H(y)}
\left[\frac{3}{2}(1-\alpha_M)\Omega_M +\frac{n_X}{2}(1-\alpha_X)\Omega_X (1+y)^{n_X-3}
\right]D_1(y,z) &  {\rm if}\ y \leq z, \\& \\
0 & {\rm elsewhere},
\end{array}\right.\nonumber
\end{eqnarray}
and the iterated kernels $K_i$ defined by the recurrence formula
\begin{equation}
\label{vol16}
K_{i+1}(y,z) \equiv \int_0 ^y K(y,x)K_i(x,z)dx .
\end{equation}

Since for all $i$, $K_i(y,z)$ and $D_1(y)$ are no negative, we see from
equations (\ref{vol14})-(\ref{vol17}) that the diameter angular distance
$D_A(z)$ is a decreasing function of both $\alpha_M$ and $\alpha_X$,
\begin{equation}
\label{vol18}
D_A(z,\alpha_M^{(1)}) \leq D_A(z,\alpha_M^{(2)}) \ {\rm for}\
\alpha_M^{(1)}
\geq
\alpha_M^{(2)},
\end{equation}
\[
D_A(z,\alpha_X^{(1)}) \leq D_A(z,\alpha_X^{(2)}) \ {\rm for}\
\alpha_X^{(1)}
\geq
\alpha_X^{(2)}.
\]

\begin{figure}
\epsfxsize=8.5cm
\centerline{\epsffile{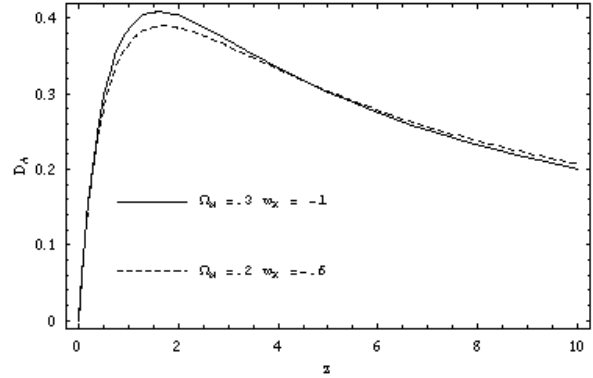}}
\caption{The angular diameter distance fortwo different FLRW universes. The unit of
distance is taken to be $c/H_0$.}
\label{dist_om_w}
\end{figure}
\begin{figure}
\epsfxsize=8.5cm
\centerline{\epsffile{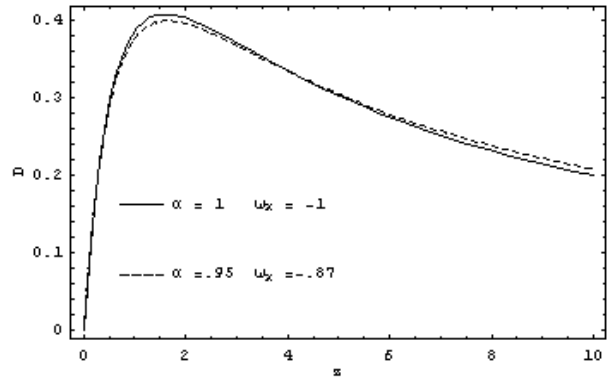}}
\caption{The angular diameter distance fordifferent values of $\alpha_M$ and $w_X$
($\Omega_M$ is equal to 0.3). The unit of distance is taken to be$c/H_0$.}
\label{dist_al_w}
\end{figure}

\section{Parameter degeneration}

As seen, the consideration of the DR equation in its full generality, with
respect to the case of a homogeneous cosmological constant, demands the
introduction of new parameters. Let us study the case of homogeneous dark
energy ($\alpha_X =1$).
\begin{figure}
\epsfxsize=13.5cm
\centerline{\epsffile{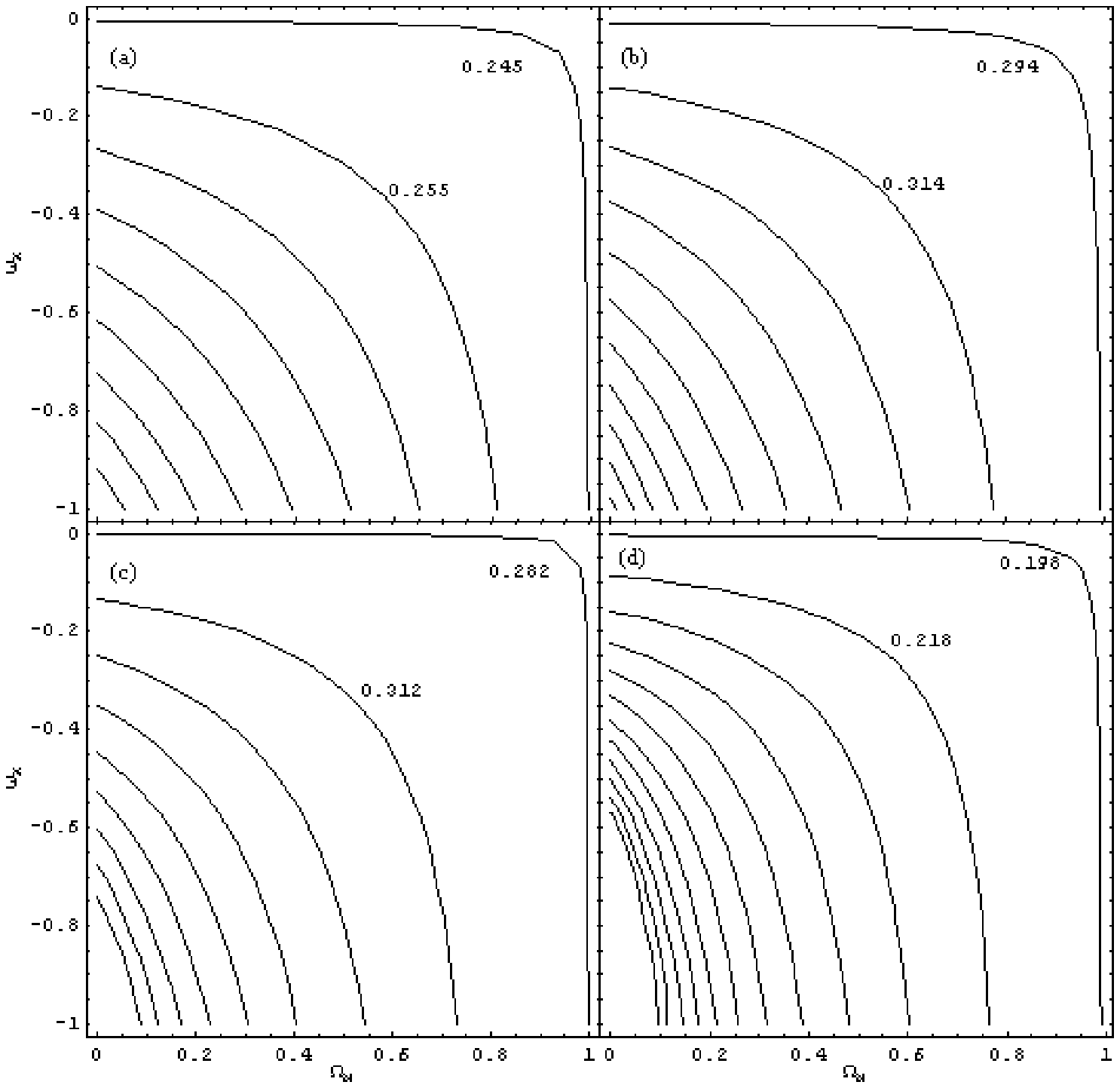}}
\caption{The angular diameter distance in the$\Omega_M - w_X$ plane, when $\alpha_M = \alpha_X= 1$.
The distance increases from the top-right to the bottom-left corner. {\em
a)} We assume $z=0.5$; each contour is drawn with steps of 0.01. {\em b)}
We assume $z=1$; the step is 0.02. {\em c)} We assume $z=2$; the step is
0.03. {\em d)} We assume $z=5$; the step is 0.02. The unit of distance is
taken to be $c/H_0$.}
\label{deg_om_w}
\end{figure}
\begin{figure}
\epsfxsize=13.5cm
\centerline{\epsffile{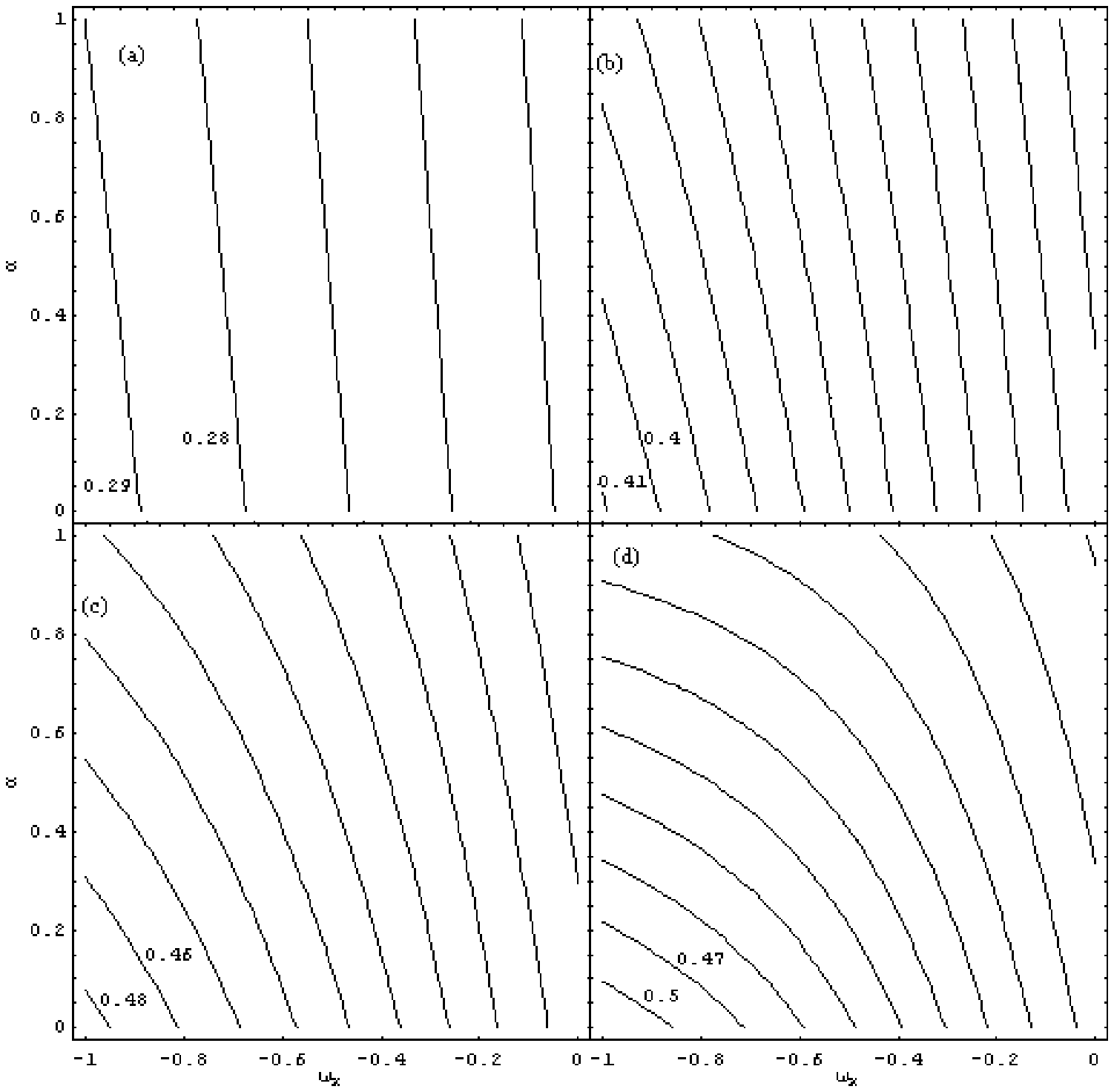}}
\caption{The angular diameter distance in the$w_X - \alpha_M$ plane, when $\Omega_M= 1$ and $\alpha_X=1$.
The distance increases from the top-right to the bottom-left corner. {\em
a)} We assume $z=0.5$; each contour is drawn with steps of 0.01. {\em b)}
We assume $z=1$; the step is 0.01. {\em c)} It is $z=2$; the step is 0.02.
{\em d)} it is $z=5$; the step is 0.03. The unit of distance is taken to
be $c/H_0$.}
\label{deg_al_w}
\end{figure}
For $\alpha_X =1$, equation (\ref{vol14}), in units of $c/H_0$, simplyfies
to
\begin{equation}
\label{deg1}
D_A(z)=D_1(z)+\int_0 ^z \sum_{i=1}^{\infty}K_i(y,z)D_1(y)dy,
\end{equation}
while equation (\ref{vol17}) reduces, for $y \leq z$, to
\begin{equation}
\label{deg2}
K_1(y,z)=\frac{3}{2}(1-\alpha_M)\Omega_M \frac{H_0}{H(y)}(1+y)^2 D_1(y,z).
\end{equation}
Some monotonical properties with respect to the cosmological parameters are
then easily derived. For an accelerated universe ($w_X <-1/3$), it is
\begin{equation}
\label{deg3}
\frac{\partial}{\partial \Omega_X}\frac{1}{H(z)}>0,\
\frac{\partial}{\partial \Omega_X} D_1(z)>0 \ {\rm if}\ w_X <-1/3,
\end{equation}
and so, for every value of the clumpiness parameter $\alpha_M$, the angular
diameter distance increases with increasing $\Omega_X$,
\begin{equation}
\frac{\partial}{\partial \Omega_X} D_A(z)>0 \ {\rm if}\ w_X <-1/3.
\end{equation}
When $w_X >-1/3$, the inequalities in equation (\ref{deg3}) are reversed
and the distance decreases with increasing $\Omega_X$. With respect to the
equation of state $w_X$, it is
\begin{equation}
\label{deg4}
\frac{\partial}{\partial w_X}\frac{1}{H(z)}<0,\
\frac{\partial}{\partial w_X} D_1(z)<0 ;
\end{equation}
and so
\begin{equation}
\label{deg5}
\frac{\partial}{\partial w_X} D_A(z)<0;
\end{equation}
large values of the distances correspond to large negative values of the
pressure of quintessence so that, for fixed $\Omega_M$, $\Omega_X$ and
$\alpha_M$, the angular diameter distance takes its maximum when the dark
energy is in the form of a cosmological constant.

We now want to stress the dependences of the angular diameter distance on
$w_X$, $\Omega_M$ and $\alpha_M$ in flat universes with $\alpha_X=1$. As
Fig.\ref{dist_om_w} and Fig.\ref{dist_al_w} show, the angular diameter
distance is degenerate with respect to different pairs of parameters, since
the distance in the $\Lambda$CDM model with
 $\Omega_M = 0.3$ is not distinguishable, within the
current experimental accuracy (Perlmutter et al. 1999), from the one in a
FLRW universe with less pressureless matter but a greater value of $w_X$ or
from an inhomogeneous universe with greater $w_X$ and the same content of
matter.

In Fig.\ref{deg_om_w}, we plot the degenerate values of the distance in
the $\Omega_M-w_X$ plane when universe is homogeneous  for four different
source redshifts: as expected, the dependence of the distance on the
cosmological parameters increases with the redshift of the source. A
general feature is that the distance is less sensitive to the components
of the universe when $\Omega_M$ is near unity and $w_X$ goes to 0. This is
easily explained: when $\Omega_M$ is large, quintessence density
$\Omega_X$ is not, and the pressureless matter characterizes almost
completely the universe; moreover, a value of $w_X$ near zero describes a
dark energy with an equation of state very similar to that of the ordinary
matter. So, increasing $w_X$ mimics a growth in $\Omega_M$. On the other
side, for low values of $\Omega_M$ ($w_X$) the distance is very sensitive
to $w_X$ ($\Omega_M$) and this effect increases with the redshift. We see
from Fig.\ref{deg_om_w} that the effects of $w_X$ and $\Omega_M$ are of
the same order for a large range of redshifts.

In Fig.\ref{deg_al_w} we compare, for $\Omega_M$ fixed to 0.3 and for
different source redshifts, the compelling effects of $\alpha_M$ and $w_X$
on the distance. When $\alpha_M$ goes away from the usually assumed value
($\alpha_M=1$), once fixed the redshift, the distance increases; on the
contrary, for $w_X$ that goes away from the value corrensponding to the
cosmological constant ($w_X
=-1$), the distance decreases. The dependence of the distance on $\alpha_M$
increases very rapidly with $z$, and, when $z=5$, the effects of $\alpha_M$
and $w_X$ are of the same order. From Fig.\ref{deg_al_w} we deduce that the
dependence on $\alpha_M$ increases when $w_X$ goes to $-1$, since values of
$w_X$ near zero have the effect to smooth the universe. In fact, when
$w_X=-1$, both a fraction $\alpha_M$ of the pressureless matter and of the
cosmological constant are uniformly distributed; when $w_X
\rightarrow 0$, quintessence behaves like ordinary matter, and so, for the
same value of $\alpha_M$, the pressureless matter homogeneously distributed
is $\alpha_M\Omega_M +\Omega_X=1-(1-\alpha_M)\Omega_M$. Intermediate values
of $w_X$ interpolate between these two extreme cases.

\begin{figure}
\epsfxsize=8cm
\centerline{\epsffile{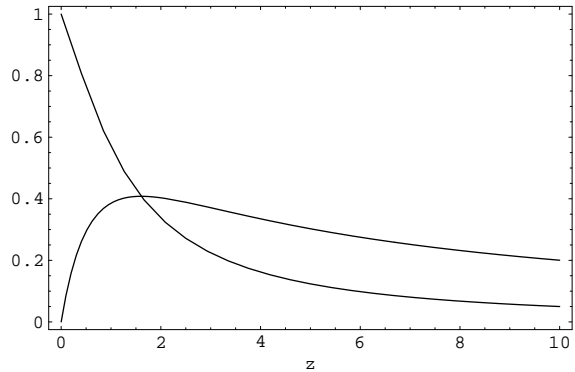}}
\caption{For flat homogeneous universes $z_m$is determined by the intercept between the angular
diameter distance and the always decreasing Hubble distance. The values on
the ordinate axis are in units of $c/H_0$. It is $\Omega_M =0.3,\
w_X=-1$.}
\label{dis-hub}
\end{figure}
\begin{figure}
\epsfxsize=8cm
\centerline{\epsffile{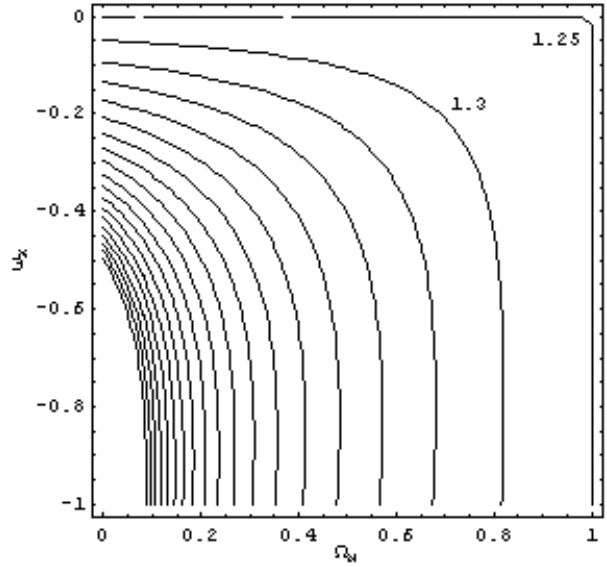}}\caption{Contours of equal $z_m$ on the $\Omega_M-w_X$
plane for flat homogeneous universes. Each contour is drawn with a step of
0.05.}
\label{zmax_om_w}
\end{figure}

\begin{figure}
\epsfxsize=9cm
\centerline{\epsffile{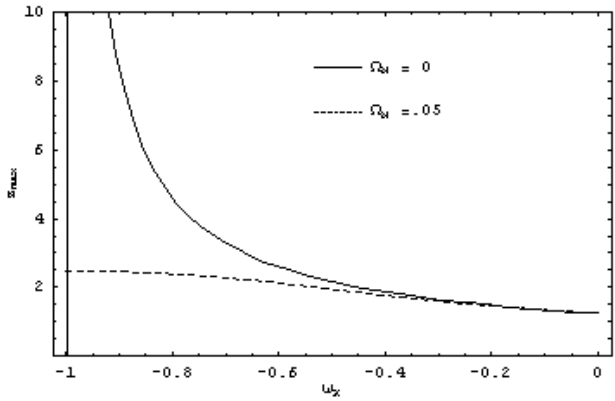}}
\caption{$z_m$ as a function of $w_X$ for two valuesof $\Omega_M$. For $\Omega_M=0.05$, $z_m$ nearly halves
itself (from 2.47 to 1.25) when $w_X$ goes from $-1$ to $0$.}
\label{zmax_w}
\end{figure}

\begin{figure}
\epsfxsize=8cm
\centerline{\epsffile{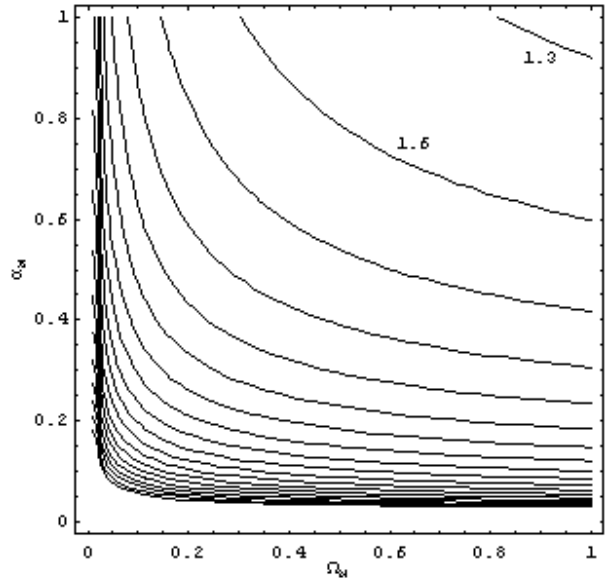}}
\caption{Contours of equal $z_m$ on the $\Omega_M-\alpha_M$plane for universes with a cosmological constant.
Contours are drawn with steps of 0.03.}
\label{zmax_om_al}
\end{figure}

\begin{figure}
\epsfxsize=8cm
\centerline{\epsffile{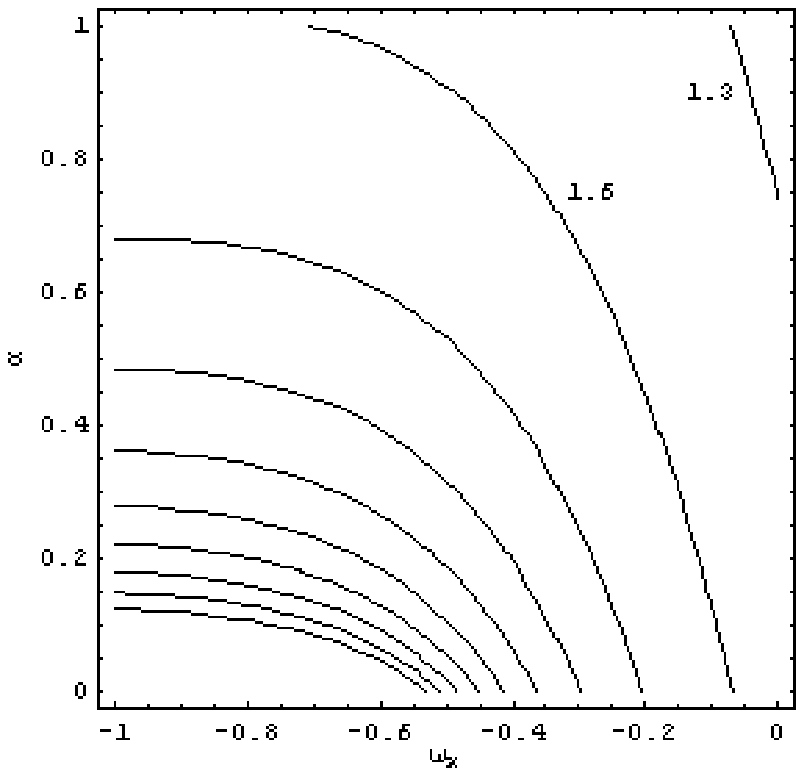}}
\caption{Contours of equal $z_m$ on the $w_X-\alpha_M$
plane for flat universes with $\Omega_M=0.3$ and $\alpha_X=1$. Contours
are drawn with steps of 0.03.}
\label{zmax_w_al}
\end{figure}

\section{The critical redshift}
The critical redshift at which the angular diameter distance of an
extragalactic source takes its maximum value has already been studied for
the case of a flat $\Lambda$CDM universe by Krauss \& Schramm
\cite{kr&sh93} and for a flat universe with quintessence by Lima \&
Alcaniz \cite{li&al00}. In this section, we will find again their results
with a new approach and will extend the analysis to inhomogeneous flat
universes.

As can be seen cancelling out the derivative of the right hand of equation
(\ref{par1}) with respect to $z$, the critical redshift $z_m$ for a flat
homogeneous universe occurs when
\begin{equation}
\label{max1}
D_A(z_d,z_m)=\frac{c}{H(z_m)},
\end{equation}
so that the angular diameter distance between an observer at $z=z_d$ and a
source at $z_m$ is equal to the Hubble distance for $z=z_m$, as you can see
in Fig.\ref{dis-hub}. Equation (\ref{max1}) is an implicit relation that
gives the dependence of $z_m$ on $z_d, \Omega_M$ and $w_X$ . Throughout
this section, we will put $z_d=0$. In Fig.\ref{zmax_om_w} we show $z_m$ for
a homogeneous flat universe. For a given value of $w_X$ ($\Omega_M$), $z_m$
decreases with increasing $\Omega_M$ ($w_X$); when $\Omega_M=0$, $z_m$
diverges for $w_X
=-1$, but also a small value of $\Omega_M$ is sufficient to eliminate this
divergence (see
Fig.\ref{zmax_w}). The minimum value of $z_m$ corresponds to the
Einstein-de Sitter universe ($\Omega_M =1$ or $w_X=0$), when $z_m=1.25$. As
you can see from Fig.\ref{zmax_om_w}, for values of $w_X$ in the range
$(-1,-0.8)$, once fixed $\Omega_M$, $z_m$ is nearly constant and this trend
increases with $\Omega_M$; on the contrary, for small $\Omega_M$
$(\stackrel{<}{\sim} 0.4)$ and $w_X
\stackrel{>}{\sim} -0.4$, $z_m$ is very sensitive to $w_X$. The small changes
 of $z_m$ in the region of large $\Omega_M$ and $w_X$ are
explained with considerations analogous to those already made in the
previous section for the values of the distance in the $\Omega_M-w_X$
plane.

Let us go now to analyse the effect of $\alpha_M$ on $z_m$. By
differentiating equation (\ref{par7}) and equation (\ref{par7bis}) with
respect to $z$, we see that the derivatives are zero only for $z
\rightarrow
\infty$: i.e., in flat universes with totally inhomogeneous quintessence or
in a generic model with cosmological constant, the critical redshift is not
finite when $\alpha_M=0$. So with respect to $z_m$, a totally clumpy
universe, indipendently of $\Omega_M$ and $w_X$, behaves like a FLRW model
completely dominated by the vacuum energy. In fact, the cosmological
constant, differently from dark energy with $w_X > -1$, does not give
contribution to the Ricci focusing and the same occurs for the pressureless
matter with $\alpha_M=0$. In Fig.\ref{zmax_om_al} we show $z_m$ in the
$\Omega_M -\alpha_M$ plane for $w_X$ fixed to $-1$. The critical redshift
decreases with increasing $\Omega_M$ and $\alpha_M$, and takes its minimum
for the Einstein-de Sitter universe $(\Omega_M =\alpha_M
=1)$, that is when the focusing is maximum. On the other side, $z_m$ is very
sensitive to $\alpha_M$, especially for large values of $\Omega_M$ since
$\alpha_M$ appears in the DR equation as a multiplicative factor of
$\Omega_M$. For $\Omega_M=0.3$, $z_m=1.61$ and 3.23 for, respectively,
$\alpha_M=1$ and 0.2, a variation of 100\%. So, combining different
cosmological tests to constrain the other cosmological parameters, we can
use the redshift-distance relation to guess the smoothness parameter
$\alpha_M$ in a quite efficient way.

We conclude this section comparing the influence of $\alpha_M$ and $w_X$ on
the critical redshift. Fig.\ref{zmax_w_al} displays $z_m$ in the
$\alpha_M-w_X$ plane, for $\Omega_M$ fixed to 0.3 and with $\alpha_X=1$. As
expected, $z_m$ increases when the focusing decreases, that is for small
values of $\alpha_M$ and $w_X$. We can see that the effects of $\alpha_M$
and $\Omega_M$ are of the same order.

\section{Conclusions}

The question of light propagation in an inhomogeneous universe is an open
topic of modern cosmology and the necessity of a positive solution grows up
with the increasing means of experimental technology that begins to explore
very high redshifts. In particular, the cosmological distances, which
provide important probes of the universe, are very sensitive to
cosmological inhomogeneities. In fact, in the present quite poor data
samples, sources usually appear dimmer and smaller with respect to
homogeneous models for the systematic effect of under-densities along the
different lines of sight. Here, we have investigated the properties of the
angular diameter distance in presence of either clumpy pressureless matter
and inhomogeneous dark energy using the 'empty beam approximation'.

The equation for the angular diameter distance with respect to the redshift
has been found in a way that is indipendent from the focusing equation. The
multiple lens-plane theory allows to derive the DR equation for
inhomogeneous quintessence, in a way that makes clear the importance of the
'empty beam approximation' in the gravitational lensing.

We have given useful forms for the distance. For non flat universe, we have
studied the case of cosmic string network, when the angular diameter
distance is expressed in terms of hypergeometric functions, and an
accelerated universe with $w_X =-2/3$ (domain walls), when the
distance-redshift relation is given in terms of Heun functions. For the
very interesting case of flat universes with inhomogeneous quintessence, we
have obtained the solution of the DR equation in terms of hypergeometric
functions. Then, we have listed the expressions for the distance when it
takes the form of elementary functions for two particular cases of flat
models: strictly homogeneous (we have considered cosmic string networks
again) and totally inhomogeneous universes, i.e. both pressureless matter
and dark energy solely in clumps.

As it could be reasonably expected, the angular diameter distance is
degenerate, also considering a smooth quintessence, with respect to
cosmological parameters: although there is a strong dependence both on the
smoothness parameter $\alpha_M$ and the equation of state $w_X$, the
intrinsic systematic errors prevent us to univocally determine the
cosmological parameters only from measurements of distance. Nevertheless,
combinig these data with indipendent observations like that concerning the
cosmic microwave backgroung radiation, we can estimate the smoothness
parameter $\alpha_M$ by determining the critical redshift, for example.

The effect of $w_X$ and $\alpha_M$, especially at high redshifts, are of
the same order but very different. An increase in $w_X$ mimics a larger
$\Omega_M$, while a more inhomogeneous universe (small $\alpha_M$) mimics
the effect of a cosmological constant. This last statement agrees with
what found by C\'{e}l\'{e}rier \cite{ce00}, in a very different context, that is
by relaxing the hypothesis of large scale homogeneity for the universe. We
have obtained a similar result by adopting the cosmological principle.

\section*{Acknowledgments}
The authors wish to thank G. Platania, C. Rubano and P. Scudellaro for
discussions and valuable comments on the manuscript.


\end{document}